# Low-Cost Implementation of Bilinear and Bicubic Image Interpolation for Real-Time Image Super-Resolution

[1]Donya Khaledyan*, [1]Abdolah Amirany, [1]Kian Jafari, [1]Mohammad Hossein Moaiyeri,
[2]Abolfazl Zargari Khuzani, [3]Najmeh Mashhadi
[1]Faculty of Electrical Engineering, Shahid Beheshti University, Tehran, Iran.
[2]Department of Electrical and Computer Engineering, University of California, Santa Cruz, USA
[3]Department of Computer Science and Engineering, University of California, Santa Cruz, USA
* d.khaledyan@mail.sbu.ac.ir

*Abstract*— Super-resolution imaging (S.R.) is a series of techniques that enhance the resolution of an imaging system, especially in surveillance cameras where simplicity and low cost are of great importance. S.R. image reconstruction can be viewed as a three-stage process: image interpolation, image registration, and fusion. Image interpolation is one of the most critical steps in the S.R. algorithms and has a significant influence on the quality of the output image. In this paper, two hardware-efficient interpolation methods are proposed for these platforms, mainly for the mobile application. Experiments and results on the synthetic and real image sequences clearly validate the performance of the proposed scheme. They indicate that the proposed approach is practically applicable to real-world applications. The algorithms are implemented in a Field Programmable Gate Array (FPGA) device using a pipelined architecture. The implementation results show the advantages of the proposed methods regarding area, performance, and output quality.

## I. INTRODUCTION

High resolution (H.R.) means pixel density within the image is distinguished, and the super-resolution (S.R.) process is one of the ways that bring us to this purpose[1]. In super-resolution to improve image quality, the size of the image will be expanding as well to improve image quality for different purposes. The importance of super-resolution algorithms in today's world to serve human beings is inexhaustible. Especially in surveillance applications, the timely execution of the algorithm is significant. In a wide range of humanitarian applications like security applications [۲-۳], face detection systems, self-driving cars, computer-aided detection systems [4-۷], and robot-assisted surgery systems, quality of the image, low cost, and real-time processes are the key points to show they are successful and pioneers.

In super-resolution, interpolation plays an essential rule and is the bottleneck of the SR algorithms. How to resize the image considering as much information as possible is an issue of concern in many applications [۸]. So, designing an optimal system for this step is important. Field Programmable Gate Array (FPGA) is a suitable platform to reach this goal. Interpolation is the process of calculating the intermediate values of a continuous event from available discrete samples. It is practiced extensively in digital image processing to magnify or reduce images and to correct spatial distortions [1, 9].

Among existing image interpolation techniques [10], nearest neighbor, bilinear, and bi-cubic interpolations have become popular. One of the causes of this prevalence is the capability of implementing these mentioned methods on the hardware platform. Due to difficulties of blocking artifacts and blurring effects in more straightforward methods such as the nearest neighbor and bilinear, the bicubic interpolation is used for superior interpolation quality because of the amount of data associated with digital images. On the other hand, the volume of computing for bicubic interpolation is high.

The bicubic interpolation algorithm addressed in this paper is a simplified computation complexity version of the algorithm presented in [11]. The proposed architecture is real-time and applicable in many security and surveillance applications, such as identifying the car plates, authentications, remote sensing [12], and similar practices, where real-time processing is necessary. A low-cost architecture of the bilinear interpolation is also proposed. The bilinear and bicubic interpolation algorithms presented here are implemented using a pipelined parallel architecture to improve the throughput for the real-time applications on FPGA. These architectures provide real-time outcomes, since after an initial latency, every pixel is estimated at the input data rate. This feature is especially important in applications such as surveillance camera.

The rest of the paper is organized as follows: In section II, the bicubic and bilinear methods are explained. In section III, some related works are discussed. Section IV presents the proposed architectures and implementation results. Finally, section V concludes the paper.

## II. BICUBIC AND BILINEAR INTERPOLATION

### A. BICUBIC INTERPOLATION

The bicubic interpolation method efforts to fit a surface among four corner pixels using a third-order polynomial function [13]. In order to compute a bicubic interpolation, the intensity values and the horizontal, vertical, and diagonal derivate at the four corner points should be calculated. The interpolated surface, $f(x,y)$, described by a third-order polynomial given by Eq. (1)

$$\sum_{i=0}^{3}\sum_{j=0}^{3} a_{ij} \times x^i y^j \quad (1)$$

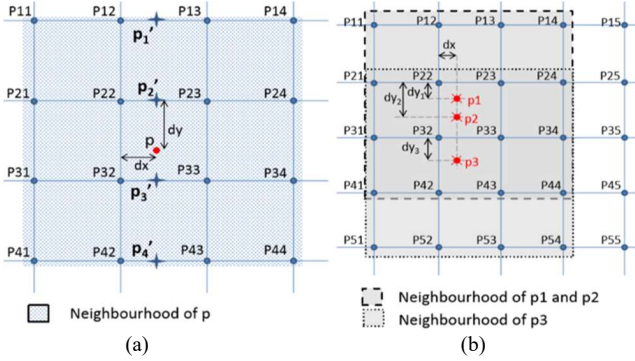

Fig. 1. Bicubic coefficients calculation (a) The neighborhood of a point P in a 2-D image space (b) Common neighborhood of points $P_1$, $P_2$, and $P_3$

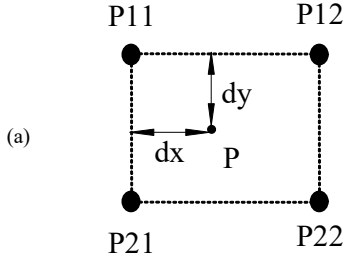

Fig. 2. The 4 neighborhood of a point 'p' in a 2-D image space

There are 16 coefficients ($a_{ij}$) that we determine to compute the function expressed by Eq. (1), each one of these 16 pixels due to their distance from the location of the reference pixel (see Fig.1) will take a coefficient.

To create the pipelined parallel architecture, first, based on Eq. (2), the interpolated pixel is calculated in each row. The results $p1'$, $p2'$, $p3'$ and $p4'$ are horizontal interpolated pixels. The final pixel will be calculated based on Eq. (3).

$$P'_i = P_{i1} \times W_{r1}(dx) + P_{i2} \times W_{r2}(dx) + P_{i3} \times W_{r3}(dx) + P_{i4} \times W_{r4}(dx) \quad i = 1.2.3.4 \quad (2)$$

$$P = P'_1 \times W_{c1}(dy) + P'_2 \times W_{c2}(dy) + P'_2 \times W_{c3}(dy) + P'_2 \times W_{c4}(dy) \quad (3)$$

Here $W_{ri}$ and $W_{ci}$ are the coefficients of the $i^{th}$ row and column, respectively. To compute these coefficients, the most current interpolation kernel is the one proposed in [14]. The same kernel function is expressed by Eq. (4). We utilize this kernel in our proposed architectures.

For hardware implementation, the most critical step is computing the coefficients of Eq. (4). If the exact values adopted in the hardware implementation, the volume of computation will be increased. Therefore, in this paper, the approximate coefficients are used to benefit from the advantages of approximate computing [15-17]. In section III it will be discussed in detail. In comparison with the bilinear interpolation, the IQS (image quality assessment) is higher, but the hardware resources are also more. The selection between these two depends on the user request.

$$w(d) = \begin{cases} \frac{3}{2}|d| - \frac{5}{2}|d|^2 + 1 & 0 \le |d| < 1 \\ \frac{-1}{2}|d|^3 + \frac{5}{2}|d|^2 - 4|d| + 2 & 0 \le |d| < 1 \\ 0 & O.W \end{cases} \quad (4)$$

### B. Bilinear Interpolation

The bilinear interpolation techniques are among the most well-known methods used in image processing due to their arithmetic simplicity [18]. It combines the values of the four nearest pixels using separable linear interpolation, as shown in Fig. 2, based on the horizontal and vertical distance from neighborhood pixels' coefficient will be calculated. The ultimate value of interpolated pixel calculated through Eq. (5).

$$P = P_{11}(1-dy)(1-dx) + P_{12}(1-dy)dx$$
$$P_{21}dy(1-dx) + P_{22}dxdy \quad (5)$$

### III. BACKGROUND

In this section, We analyze and investigate several hardware implementations of bicubic and bilinear interpolation.

In [18], a Real-time FPGA Implementation of Barrel distortion correction method by using bilinear interpolation is presented. The architecture in [11] grants high output quality but demands very high output resources; hence consumes high power.

In [11, 19], the bicubic interpolation is implemented. These architectures store the entire image pixels in external memory. Hence, sizable external memory is required. This external memory increases the overall cost of the system, reduces the performance, and increases the power consumption.

In [20], a comparison between classical interpolation and new convolution-based interpolation is presented. This comparison includes other cubic interpolation systems not earlier studied in signal and image processing. The experimental results in [20] also compare the computational complexity of these methods.

Most of the bicubic interpolation implementations using FPGA for image scaling [19, 21] typically use floating-point units (FPU). The FPU imposes a significant area overhead, consumes high power, and affects the overall performance of the system. In [19], a lookup table method, along with parameterized modules, is used instead of a floating-point multiplayer.

In [22], a different interpolation kernel is established based on five independent parameters that measure its angular frequency, amplitude, standard deviation, and duration. However, this method has complexity in computation and is not suitable for hardware implementation. To overcome this computational complexity, a novel low-complexity cubic interpolation implementation for spaceborne georeferencing images is proposed in [23]. While the architecture in [23] reduces the computational complexity, it is only applicable to the spaceborne georeferencing images.

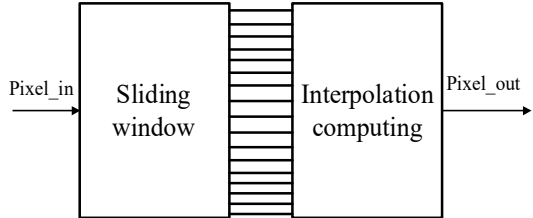

Fig. 3. The block diagram of the proposed algorithm

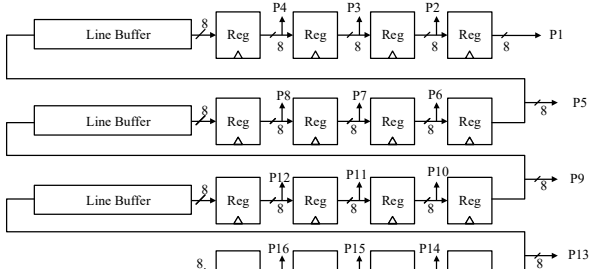

Fig. 4. The block diagram for 4*4 sliding window

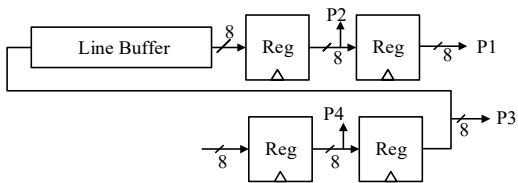

Fig. 5. The block diagram for 2*2 sliding window

## IV. PROPOSED ARCHITECTURES FOR BICUBIC AND BILINEAR INTERPOLATION AND IMPLEMENTATION RESULTS

### A. Proposed Architectures

Proposed architecture for bilinear and bicubic interpolation consists of 2 main steps, are shown in Fig. 3.

First, the architecture provides proper pixels for interpolation computing part. And then determining the coefficients and calculate the interpolated pixels. It is clear that step 2 is more critical, and the main idea of this paper is in this step.

The architecture presented in [11, 19] stores all of the image pixels in external memory. However, in our proposed architecture, by using the sliding window, which is presented in detail in Fig. 4, there is no need to save the whole image. Thus, the first step provides a good saving in memory and, as a result, in hardware resources. The size of the line buffer is equal to the length of the image. As in the bicubic interpolation, 16 neighborhoods require to be read; The sliding window has 3 line buffers. Fig. 4 shows the architecture of the sliding-window for bicubic interpolation.

Accordingly, as four neighborhood pixels are required for the bilinear process, we just need a line buffer. Fig .5 shows the architecture of the sliding-window for bilinear interpolation.

After the first step and providing the pixels for the second step, the pixels are multiplied by the coefficients. If the exact values are used, a large number of multipliers will be needed. In this paper no multiplier block is used for interpolation implementation. However, as this is a trade-off between hardware resources and accuracy, we have developed an approximated bicubic- and bilinear-based method, which is more suitable for computational systems with limited memory, such as FPGAs and DSPs [24]. The block diagram of the interpolate part for bilinear, and bicubic interpolations are shown in Figs. 6 and 7, respectively.

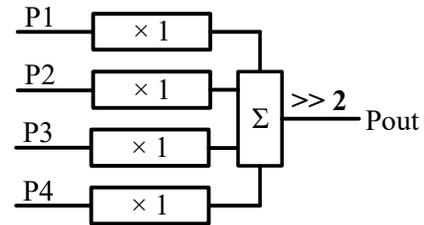

Fig. 6. The Architecture of final value calculation of bilinear interpolation

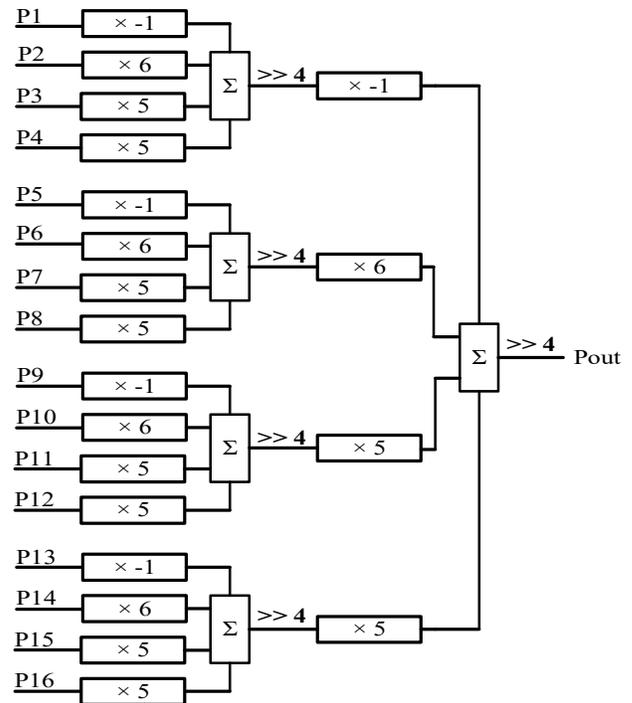

Fig. 7. The Architecture of final value calculation of bicubic interpolation

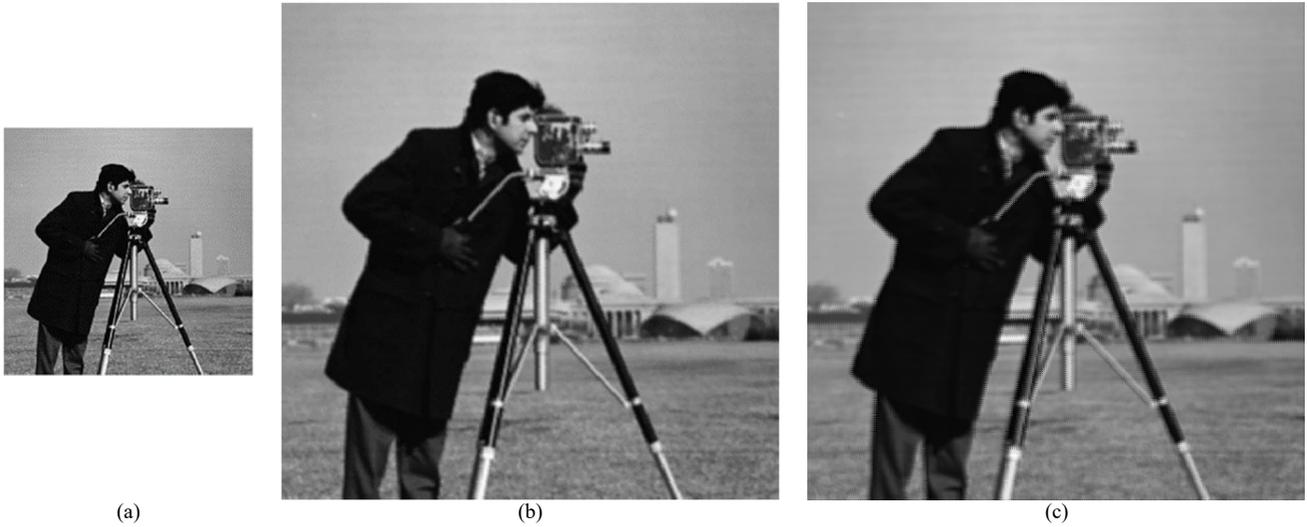

(a)          (b)          (c)

Fig. 8. The final results (a) Input image (b) Result of the bilinear interpolation (c) Result of the bicubic interpolation

TABLE I. COMPARISON OF THE PROPOSED ARCHITECTURE WITH RECENT INTERPOLATION METHODS

| Architectures | Image size | Interpolation algorithm | Implementation platform | Frequency (Mhz) | Slice LUTs | Slice registers | Block RAM | DSP |
|---|---|---|---|---|---|---|---|---|
| Proposed in [14] | 640*480 | Linear | Virtex-2 | 104.3 | NA | NA | NA | NA |
| Proposed in [25] | 2560*1920 | Cubic | Virtex-6 | 130.0 | NA | NA | NA | NA |
| Proposed in [11] | 2560*1920 | Cubic | Virtex-6 | 75.0 | 7900 | 7843 | 78 | 48 |
| Proposed in [23] | 256*256 | Cubic | Artix-7 | 100.0 | 5293 | 8432 | 102 | 39 |
| Bi-linear | 256*256 | Linear | Artix-7 | 314.8 | 97 | 44 | 0 | 0 |
| Bicubic | 256*256 | Cubic | Artix-7 | 289.2 | 359 | 162 | 0 | 0 |

TABLE II. PSNR AND SSIM OF OUTPUT IN COMPARISON TO THE SOFTWARE METHOD

| Images | Bi-linear | | Bicubic | |
|---|---|---|---|---|
| | PSNR | SSIM | PSNR | SSIM |
| Cameraman | 23.77 | 0.972 | 29.02 | 0.987 |
| Moon | 25.76 | 0.961 | 29.92 | 0.974 |
| Rice | 23.77 | 0.973 | 29.19 | 0.985 |
| Coins | 24.06 | 0.974 | 28.42 | 0.990 |

## B. Implementation and Simulation Results

The implementation and simulation of the proposed architecture are done using the ISE design suite and MATLAB.

As mentioned before, the hardware-based design techniques such as parallelism and pipelining techniques can be developed on an FPGA, which is impossible in a dedicated DSP design. FPGA is a matrix of logic blocks that are combined by a network of switches. Logic blocks and switching networks are reconfigurable, allow in application-specific hardware to be formed. As FPGA allows a compromise among the adaptability of general-purpose processors and the hardware-based speed of ASICs.

By implementing image processing algorithms on reconfigurable hardware, the time to market costs can be reduced. Besides, it can enable quick prototyping of complicated algorithms, and simplifies the debugging and verification phases. So, FPGAs are reliable options for the implementation of real-time image processing algorithms. The advantage of the FPGA-based interpolation is that the design can be implemented in smart camera designs, which means, it is useable in embedded systems where the sensor is attached to the FPGA for pixel data processing. These kinds of applications usually produce low-cost and real-time processing devices.

Figure 8 shows the results of the proposed architectures. As indicated in Fig. 8, the proposed architectures provide an acceptable output quality while occupying low resources and delivers high performance.

Table I shows the results of the implementation of the proposed architectures. As this table exhibits, thanks to the approximate coefficients and pipelined architecture, the proposed architectures offer high frequency and occupy low resources with negligible lower output quality. Table II gives the peak signal to noise ratio (PSNR) [26] and structural similarity (SSIM) [27] of the proposed architectures for different input images. As this table shows, the proposed methods offer high PSNR and SSIM. With reducing the image dimensions, the PSNR and SSIM will be reduced as well. Therefore, medium size images are selected in this paper to have pessimistic results. However, by this choice, the hardware resources are reduced. Notably, even if we utilize large size images, the hardware resources are much less than the other works like [11, 14, 22, 24].

As proposed architectures offer high frequency, low power consumption, and occupy small resources with negligible lower output quality, we can utilize it in applications such as surveillance cameras where hardware resources and power are limited. Besides, according to the mentioned features (high frequency, low power consumption, low area overhead, and negligible lower output quality), the proposed architectures can be implemented in applications such as internet of things (IoT) nodes and sensors, communication and information technologies, and mobile clinics which are also facing the mentioned limitations.

V. CONCLUSION

Interpolation is one of the critical steps in super-resolution techniques, tracking systems, robotic, online videos, mobile applications, and most importantly, security applications like surveillance cameras. In this paper, an FPGA implementation for the improved bicubic and bilinear convolution interpolation for real-time applications is proposed. The proposed method reduces the computational complexity, enhances the speed, and reduces the FPGA resources while providing an excellent trade-off between image quality and calculation simplicity. Due to the few computational requirements and real-time capability of the proposed architecture, it can be considered a reasonable solution for applications that require interpolation in real-time with the minimum cost in hardware.